\documentclass[conference]{IEEEtran}
\IEEEoverridecommandlockouts

\usepackage{cite}
\usepackage{amsmath,amssymb}
\usepackage{graphicx}
\usepackage{booktabs}
\usepackage{cuted}   
\usepackage{capt-of} 
\usepackage{url}
\usepackage{pifont}  
\usepackage{tikz}
\usetikzlibrary{arrows.meta,positioning,fit,calc,backgrounds}

\graphicspath{{figs/}}

\newcommand{\cmark}{\ding{51}}
\newcommand{\xmark}{\ding{55}}
\newcommand{\pmark}{$\sim$}      

\begin{document}

\title{Fluxion: A Differentiable DSP Runtime from Training GPU to Real-Time Edge}

\author{\IEEEauthorblockN{Matteo Spanio}
\IEEEauthorblockA{\textit{Centro di Sonologia Computazionale (CSC)} \\
\textit{Dept.\ of Information Engineering, University of Padova}\\
Padova, Italy \\
spanio@dei.unipd.it}
\and
\IEEEauthorblockN{Antonio Rod\`a}
\IEEEauthorblockA{\textit{Centro di Sonologia Computazionale (CSC)} \\
\textit{Dept.\ of Information Engineering, University of Padova}\\
Padova, Italy \\
roda@dei.unipd.it}
}

\maketitle

\begin{strip}
\vspace{-1.2em}
\centering
\resizebox{0.9\textwidth}{!}{
\begin{tikzpicture}[
  font=\footnotesize,
  box/.style={draw, rounded corners=3pt, align=center, inner sep=5pt, line width=0.9pt},
  owned/.style={box, fill=oiBlue!14, draw=oiBlue!70!black},
  gate/.style={box, fill=oiAmber!22, draw=oiAmber!60!black},
  edge/.style={box, fill=oiGreen!14, draw=oiGreen!60!black},
  rented/.style={box, dashed, fill=white, draw=black!60},
  phase/.style={font=\scriptsize\scshape, text=black!45},
  metric/.style={font=\scriptsize\itshape, text=black!60, align=center, anchor=north},
  lbl/.style={font=\scriptsize\itshape, align=center, text=black!70},
  arr/.style={-{Stealth[length=2.6mm, width=2mm]}, line width=1.1pt, draw=black!70},
]
\definecolor{oiBlue}{RGB}{86,180,233}
\definecolor{oiAmber}{RGB}{230,159,0}
\definecolor{oiGreen}{RGB}{0,158,115}

\node[phase] at (2.4, 2.5)   {compose};
\node[phase] at (7.8, 2.5)   {train};
\node[phase] at (12.5, 2.5)  {certify \& freeze};
\node[phase] at (16.55, 2.5) {deploy \& play};
\foreach \x in {4.85, 10.4, 14.45}
  \draw[black!18, line width=0.6pt] (\x, 2.35) -- (\x, -2.55);

\node[owned, minimum width=42mm, minimum height=11mm] (algebra) at (2.4, 1.75)
  {\textbf{graph algebra}\\[1pt]
   $g=\mathrm{lp}(f_c)\,\big|\,(\mathrm{eq}+\mathrm{delay})$};
\node[owned, minimum width=42mm, minimum height=14mm] (numerics) at (2.4, -0.35)
  {\textbf{owned numerics}\\[1pt]
   closed-form design (\texttt{f64})\\
   kernels $\cdot$ \textbf{analytic adjoints}\\
   \textbf{stability certificates}};
\draw[arr] (algebra) -- (numerics);

\node[owned, minimum width=46mm, minimum height=13mm] (train) at (7.8, 0.95)
  {\textbf{training engine}\\[1pt]
   differentiable end-to-end\\
   in-loop stability projection};
\node[rented, minimum width=46mm, minimum height=10mm] (rentedbox) at (7.8, -1.05)
  {\emph{rented:} autodiff graph (Burn/torch/JAX)\\
   GPU codegen (CubeCL) $\cdot$ arrays (DLPack)};
\draw[dashed, -{Stealth[length=2mm]}, black!60] (rentedbox) -- (train);
\draw[arr] (numerics.east) -- ++(0.75,0) |- (train.west);

\node[gate, minimum width=18mm, minimum height=13mm] (freeze) at (11.55, 0.95)
  {\textbf{certify}\\$\downarrow$\\\textbf{freeze}};
\node[gate, minimum width=13mm, minimum height=10mm] (fxg) at (13.5, 0.95)
  {\texttt{.fxg}\\v1\;\cmark};
\draw[arr] (train) -- (freeze);
\draw[arr] (freeze) -- (fxg);

\node[edge, minimum width=35mm, minimum height=16mm] (rt) at (16.55, 0.95)
  {\textbf{real-time engine}\\[1pt]
   allocation-free $\cdot$ lock-free\\
   coefficient crossfade\\
   \emph{re-certifies on device}};
\draw[arr] (fxg) -- node[lbl, above=0.6mm]{push} (rt);

\node[metric] at (7.8, -1.85)
  {2990\,Msamples/s GPU forward (resident)\\0.0014\,dB EQ inversion};
\node[metric] at (12.5, -1.85)
  {stability verdict\\(poles $\cdot$ small gain)};
\node[metric] at (16.55, -1.85)
  {27.5\,$\mu$s on-device certify + enqueue\\2.8\,MB binary $\cdot$ 217$\times$ RT (Pi\,5)};

\node[lbl, anchor=north] at (2.4, -1.85)
  {\tikz{\node[owned, minimum width=5mm, minimum height=2.8mm, inner sep=0]{};}\;owned
   \;\;\tikz{\node[rented, minimum width=5mm, minimum height=2.8mm, inner sep=0]{};}\;rented\\[2pt]
   \tikz{\node[gate, minimum width=5mm, minimum height=2.8mm, inner sep=0]{};}\;deployment gate};
\end{tikzpicture}
}
\par\vspace{0.6em}

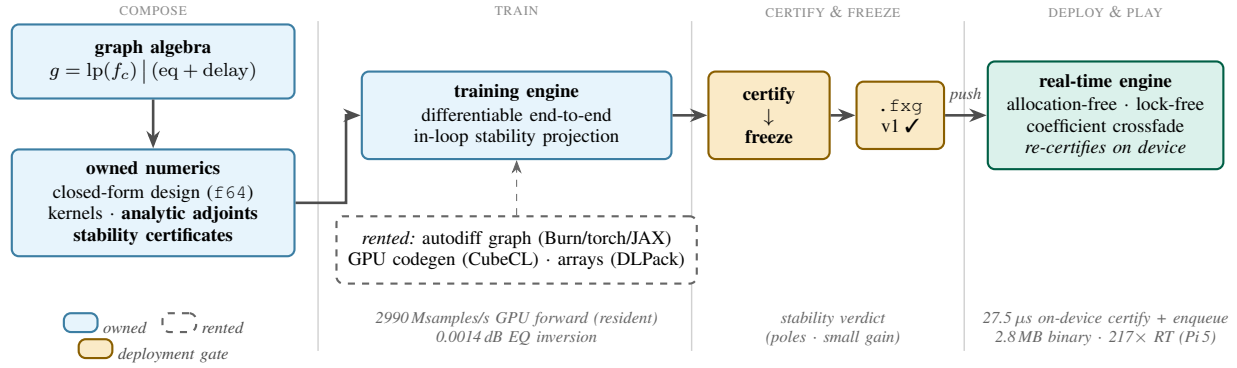
\captionof{figure}{Architecture overview. A graph algebra over
  Fluxion's own numerical code is lowered to a training engine and a
  real-time engine; the autodiff graph, GPU code generation, and array
  interchange come from existing libraries. A trained graph is checked
  for stability and frozen into a versioned artifact, which a
  Raspberry~Pi-class device checks again before crossfading it in.}
\label{fig:arch}
\end{strip}

\begin{abstract}
Native real-time audio engines meet audio deadlines but are hard to
connect to machine-learning workflows, while tensor frameworks
integrate with machine learning but process audio as offline batches
without timing guarantees; runtime layers built on them inherit the
host's scheduling and deployment footprint. This paper presents
Fluxion, a Rust audio DSP runtime for cascaded second-order sections,
FIR filters, and delay-line effects. Fluxion implements the parts that
are specific to DSP: closed-form filter design, forward kernels,
analytic adjoints for the recursion, and stability checks. It relies on
existing libraries for the autodiff graph, GPU code generation (tested
here on CUDA), and array interchange. One graph algebra is lowered to a
differentiable training engine and to an allocation-free real-time
engine; between them, a trained graph is checked for stability and
frozen into a versioned artifact that an edge device checks again
before playback. We compare Fluxion with six framework-hosted and
native baselines under one float32 protocol. Among the systems that
compute the exact recursion, Fluxion has the highest GPU throughput
when data stays on the device and the highest throughput on most CPU
workloads, while its CPU training step is slower than those of
torchaudio and GRAFX. Its test suite checks the audio thread for
allocations with an instrumented allocator, and we report measured
latency tails. Fluxion is open source at
\url{https://github.com/matteospanio/fluxion}.
\end{abstract}


\section{Introduction}
\label{sec:intro}

Native real-time engines such as Faust~\cite{orlarey2009faust},
SuperCollider~\cite{mccartney2002supercollider}, and JUCE~\cite{juce}
run audio with low latency, but they are hard to connect to the
machine-learning tools now common in audio research. Tensor frameworks
such as torchaudio~\cite{Yang2022Torchaudio,torch2},
nnAudio~\cite{cheuk_nnaudio_2020}, and GRAFX~\cite{lee_grafx_2024}
work well with machine learning, but they treat audio as offline
batches of tensors, with no streaming state and no latency bound.
Trained tensor graphs can be deployed on edge devices through TFLite
or ONNX Runtime, but a sample-recurrent filter reaches those runtimes
only by unrolling.

TorchFX~\cite{spanio2025torchfx}, our earlier system, and
differentiable-DSP libraries such as FLAMO~\cite{flamo} and
ADAC~\cite{adac} build a runtime layer on top of a tensor framework.
TorchFX, for example, records composition as a deferred graph, fuses
second-order-section (SOS) cascades into single dispatches, and reuses
the host's autograd. These layers remain limited by their host: the
interpreter schedules the real-time path, so the latency tail can be
measured but not bounded, and the Python and PyTorch stack weighs
gigabytes, which makes edge deployment impractical.

We argue for a different division of work. A DSP runtime should
\emph{own} the parts that are specific to DSP: closed-form coefficient
design, forward kernels, analytic adjoints for recursive filters, and
the stability checks that decide whether a trained filter can be
deployed. It should \emph{rent} what general-purpose ML infrastructure
already provides: the autodiff graph (Burn~\cite{burn}, PyTorch, or
JAX), GPU code generation (CubeCL~\cite{cubecl}), and zero-copy array
interchange (DLPack~\cite{dlpack}). Keeping the adjoints in the runtime
lets several autodiff frameworks use them, and relying on existing
libraries for the rest keeps the runtime small.

We present Fluxion, a Rust audio DSP runtime built this way
(Fig.~\ref{fig:arch}). A functional graph algebra (series
\texttt{|}, parallel \texttt{+}, feedback \texttt{\textasciitilde},
and named nodes) is lowered to two engines: a training engine that
composes analytic vector-Jacobian products (VJPs), and a lock-free
real-time engine that does not allocate on the audio thread. Feedback
is not yet lowered to either engine, so our results cover cascades,
FIRs, and delay-line effects. Filters are designed in double
precision; the single-precision coefficients that run are
checked for stability and frozen into a versioned artifact, which the
real-time engine plays.

\medskip\noindent\textbf{Contributions.}
\begin{enumerate}
  \item A runtime architecture in which one graph IR is lowered to a
    training engine and an allocation-free real-time engine, with an
    explicit freeze step between them (Sec.~\ref{sec:arch}).
  \item Hand-derived adjoints for the filter recursion, registered as
    custom operators in existing autodiff frameworks and extended to
    \emph{design parameters} (cutoff, gain, Q), with a stability
    projection that keeps raw-coefficient training inside the stable
    region (Sec.~\ref{sec:arch:diff}, Sec.~\ref{sec:eval-diff}).
  \item A deployment pipeline that designs filters in \texttt{f64},
    checks the stability of the frozen \texttt{f32} coefficients (pole
    radius per section, small gain for feedback loops), and stores them
    in a versioned artifact that the real-time engine loads without a
    design step (Sec.~\ref{sec:arch:freeze}).
  \item A lock-free real-time engine with coefficient crossfades, whose
    tested audio-thread paths make no allocations under a counting
    global allocator, with a control test that confirms the counter
    works (Sec.~\ref{sec:arch:rt}, Sec.~\ref{sec:eval-rt}).
  \item A float32 comparison with exact-recurrence and
    frequency-sampling baselines on throughput, training cost, latency,
    and footprint, and an industrial case study
    (Sec.~\ref{sec:eval-thr} to~\ref{sec:eval-rt}).
\end{enumerate}

\section{Related Work}
\label{sec:related}

\medskip\noindent\textbf{Tensor-framework audio.}
torchaudio~\cite{Yang2022Torchaudio,torch2} provides exact time-domain
recursion over batched tensors but no fused-cascade primitive, which
Sec.~\ref{sec:eval-thr} measures. nnAudio~\cite{cheuk_nnaudio_2020}
provides trainable GPU spectral front-ends without recursive
filtering, so we do not use it as a comparator.
GRAFX~\cite{lee_grafx_2024} composes differentiable audio graphs on
PyTorch with an exact and an approximate filtering backend, and we
measure both. Engel et al.~\cite{jesse_ddsp_2020} established the idea
of learning through signal-processing structure (DDSP). All of these
systems inherit their host's execution model.

\medskip\noindent\textbf{Differentiable DSP.}
IIR filters can be made differentiable in two main ways. One is to
build modules inside a host framework and use its autograd, as in
FLAMO's frequency-sampled modules~\cite{flamo} and ADAC's
FLAMO-to-FAUST compilation pipeline~\cite{adac}. The other is to
derive the adjoint of the recursion analytically and register it as a
custom operator. This has been done for linear dynamical
operators~\cite{forgione2021dynonet}, all-pole
filters~\cite{yu_diffapf}, and direct-form
filters~\cite{yu2025accelerating}, and torchaudio's \texttt{lfilter}
and TorchFX's opt-in \texttt{torchfx.ddsp}
module~\cite{spanio2025torchfx} also ship analytic backward passes.
Fluxion takes the second approach and keeps the adjoints in the same
runtime that checks and deploys the trained filter.
Tab.~\ref{tab:trainthr} compares the cost of both approaches.

Filters trained in FLAMO or in TorchFX's differentiable module can be
imported into Fluxion's freeze pipeline (Sec.~\ref{sec:arch:freeze}).
ADAC is a compiler without a filtering runtime of its own, and its
published benchmark measures code-generation latency, so we discuss it
without measuring it and measure FLAMO, on which it builds. ADAC's
loop-aware stability certificate is the closest prior work to our
stability verdicts; Fluxion also applies a stability projection during
training (Sec.~\ref{sec:arch:diff}).

\medskip\noindent\textbf{Native real-time engines.}
Faust~\cite{orlarey2009faust},
SuperCollider~\cite{mccartney2002supercollider}, and JUCE~\cite{juce}
run audio with low latency and complement Fluxion; Faust can also be
used for training through its JAX backend~\cite{dawdreamer}.
RTNeural~\cite{rtneural}, the Neutone SDK~\cite{mitcheltree2025neutone},
and anira~\cite{ackvaschulz2024anira} follow a two-step workflow in
which models are trained in Python and then run by a native inference
engine. anira is the closest to our setting. It runs ONNX Runtime,
LibTorch, and TensorFlow Lite models from an audio callback by moving
inference to a thread pool. By default this adds one buffer of
latency, and an optional bounded wait in the callback removes it.
Fluxion computes the DSP on the audio thread (Sec.~\ref{sec:eval-rt}
measures both approaches on the same workload) and uses one runtime
for training and deployment.

\medskip\noindent\textbf{GPU DSP.}
Parallel IIR filtering on GPUs builds on prefix
scans~\cite{blelloch1990prefix,sengupta2007scan,harris2007parallelprefix},
which TorchFX's CUDA path uses. Fluxion's GPU kernel runs each batch
row's cascade sequentially in time, with one GPU thread per row.
gpuRIR~\cite{Diaz-Guerra20215653} and Brian
Hears~\cite{fontaine_brian_2011} vectorise related audio workloads.
Burn~\cite{burn}, CubeCL~\cite{cubecl}, and DLPack~\cite{dlpack} are
dependencies of Fluxion. Tab.~\ref{tab:features} compares the systems
discussed in this section.

\begin{table*}[t]
  \caption{Feature comparison. \cmark{} supported, \xmark{} not
    supported, \pmark{} partial. ``Analytic $\nabla$'': hand-derived
    adjoints instead of unrolling on the host tape. ``Portable GPU'':
    not tied to one vendor, with validation status inline. ``RT
    verif.'': audio path checked under allocator instrumentation.
    ``Ckpt import'': reads filter checkpoints from other frameworks.}
  \label{tab:features}
  \centering
  \scriptsize
  \setlength{\tabcolsep}{3pt}
  \begin{tabular}{@{}l ccccccccc@{}}
    \toprule
    Framework & GPU & Portable GPU & Analytic $\nabla$ & Design-param $\nabla$
      & Stability cert. & Real-time & RT verif. & Edge footprint & Ckpt import \\
    \midrule
    \textbf{Fluxion} & \cmark & \pmark\,(CUDA val.) & \cmark & \cmark
      & \cmark\,(+ in-loop) & \cmark & \cmark & \cmark\,($O$(MB))
      & \cmark\,(FLAMO, TorchFX) \\
    TorchFX~\cite{spanio2025torchfx} & \cmark & \xmark & \pmark\,(opt-in)
      & \pmark & \xmark & \pmark\,(CPU meas.) & \xmark & \xmark\,($O$(GB)) & \xmark \\
    torchaudio~\cite{Yang2022Torchaudio,torch2} & \cmark & \pmark\,(PyTorch)
      & \pmark\,(\texttt{lfilter}) & \pmark & \xmark & \xmark & \xmark
      & \xmark & \xmark \\
    GRAFX~\cite{lee_grafx_2024} & \cmark & \pmark\,(PyTorch) & \pmark & \pmark
      & \xmark & \xmark & \xmark & \xmark & \xmark \\
    FLAMO/ADAC~\cite{flamo,adac} & \cmark & \pmark\,(PyTorch) & \xmark & \cmark
      & \pmark\,(design) & \pmark\,(FAUST exp.) & \xmark & \pmark
      & \pmark\,(FLAMO$\to$FAUST) \\
    anira~\cite{ackvaschulz2024anira} & \xmark & --- & \xmark & \xmark
      & \xmark & \cmark\,(default +1 buf.) & \pmark\,(sanitizer)
      & \pmark\,($O$(100\,MB)) & \cmark\,(ONNX/TS/TFLite) \\
    Faust~\cite{orlarey2009faust} & \xmark & --- & \xmark & \pmark\,(JAX)
      & \xmark & \cmark & \pmark & \cmark & --- \\
    SciPy~\cite{2020SciPy-NMeth} & \xmark & --- & \xmark & \xmark
      & \xmark & \xmark & \xmark & \pmark & --- \\
    \bottomrule
  \end{tabular}\\[2pt]
  {\footnotesize GRAFX cells refer to grafx~0.6.0. torchaudio's
   \texttt{lfilter} has its own differentiable path and its
   \texttt{functional} helpers take cutoff and Q under autograd, hence
   its two \pmark{} gradient cells. Fluxion's import cell is tested in
   CI (Sec.~\ref{sec:arch:freeze}) and covers SISO cascades only.}
\end{table*}

\section{Architecture}
\label{sec:arch}

\subsection{Graph algebra and engines}
\label{sec:arch:overview}
Users describe processing as a functional expression, in the style of
Faust's block-diagram algebra~\cite{orlarey2009faust}. Filters and
effects are values, \texttt{|} chains them in series, \texttt{+} runs
them in parallel and sums the outputs, \texttt{\textasciitilde} closes
a feedback loop, and any node can be named. The expression is stored
as a small typed graph with self-describing parameters, so Fluxion can
validate it when it is built, serialise it, and lower it to different
engines.

Fluxion has two engines because training and real-time audio have
different requirements. A training step benefits from large batches,
free allocation, and often a GPU. An audio callback needs a bounded,
deterministic code path, and our earlier measurements showed that
per-block GPU dispatch raises throughput but does not lower
latency~\cite{spanio2025torchfx}. Fluxion therefore keeps one engine
per workload and connects them through a freeze step
(Sec.~\ref{sec:arch:freeze}). The sample rate is bound when a graph is
lowered, so the same graph can run at any rate.

Both engines use the same traversal: a single walk interprets the
graph against a backend trait with one method per kernel, so the
backend supplies the buffer type and the kernels and the walk supplies
the composition. For training the buffer is a differentiable tensor;
for real time, a pre-allocated block of samples. Each lowering rejects the
operators it cannot support. The training path rejects operators
without an adjoint, and the real-time path rejects operators that need
more than the current block, such as peak normalisation. Both
currently reject feedback, so the engines cover cascades, FIRs, and
delay-line effects.

\subsection{Analytic adjoints}
\label{sec:arch:diff}
Recursive filters are a poor fit for generic automatic
differentiation. If a framework traces the recurrence sample by
sample, the tape grows with signal length and cascade depth, and Yu et
al.\ report order-of-magnitude slowdowns for this
reason~\cite{yu_diffapf}. Fluxion gives each operator a hand-derived
adjoint, that is, a closed-form backward pass. A filter section
computes
\begin{equation}
\label{eq:biquad}
	y_t = \sum_{i=0}^{M} b_i x_{t-i} - \sum_{j=1}^{N} a_j y_{t-j}.
\end{equation}
In the backward pass of Eq.~\eqref{eq:biquad}, the output cotangent
goes through a recurrence of the same form, run backwards in time.
Input gradients follow by filtering that cotangent with the numerator
taps, and coefficient gradients are correlations with the delayed
input and output signals. Design-parameter gradients follow by the
chain rule through the design map. The design map is closed-form
algebra without recursion, so Fluxion computes its Jacobian by central
differences, which costs a few extra design evaluations per step and
does not grow with signal length.

Fluxion registers these vector-Jacobian products in the autodiff
framework the user already works with, as a Burn custom operator or as
a custom autodiff operator in PyTorch or JAX. Each operator enters the
host graph as one node: forward is the ordinary kernel, backward the
hand-derived adjoint, so the host never records the recurrence. The
executor that runs a graph on the CPU is
the same walk that runs it over autodiff tensors, so series
composition chains the adjoints and parallel composition sums the
cotangents. Gradients are checked per operator and per composed graph
(Sec.~\ref{sec:eval-diff}).

Training on design parameters is common in differentiable DSP, where
cookbook design equations map cutoff, gain, and Q to
coefficients~\cite{nercessian2020neural,steinmetz2021automatic}. In
Fluxion these parameters sit on top of the analytic adjoint instead of
an unrolled tape, so a step costs one forward and one backward pass of
the recursion (Tab.~\ref{tab:trainthr}). Within the allowed parameter
ranges, every set of design parameters gives a stable filter. When raw
coefficients are trained instead, a projection moves each optimiser
step back into the stable region (Sec.~\ref{sec:arch:freeze}).

\subsection{Stability check and freezing}
\label{sec:arch:freeze}
Filters are designed in double precision and then cast to the
single-precision coefficients that run; we call this cast
\emph{freezing}. The stability \emph{certificate} is computed from the
frozen values, not from the double-precision design. For an SOS
cascade, each section is normalised to $a_0{=}1$ and the largest pole
radius after the cast is recorded. Graphs with explicit feedback also
get a small-gain check, the maximum loop gain over a linear frequency
grid on $[0, \pi]$. The shipped echo and reverb nodes have
frequency-flat loop gains, so for them the check is exact; for other
loops it is a sampled estimate. The general \texttt{\textasciitilde}
operator is checked the same way but is not yet lowered to either
engine.

The check returns one of three verdicts: \texttt{Stable} (all poles
strictly inside the unit circle), \texttt{Marginal} (within
$\pm10^{-4}$ of the boundary, accepted but flagged), or
\texttt{Unstable} (rejected). We chose the band with the cast in mind.
Half-ulp rounding of a float32 coefficient is a relative
$6{\times}10^{-8}$, and near the unit circle a pole radius moves by
about the same amount, so the band is more than three orders of
magnitude wider than the rounding error. The frozen graph is stored
in a \texttt{.fxg} file, a JSON document with a format version and a
kind tag, so a file with an unknown version is rejected instead of
misread. A \emph{graph} file stores topology and design parameters and
is sample-rate independent; a \emph{frozen-sos} file stores the float32
coefficients and their design rate, so the real-time engine loads it
without a design step. The certificate
depends only on the file contents, so a device can recompute it at
load time instead of trusting the tool that produced the file
(Sec.~\ref{sec:eval-rt}). It covers stability only and does not
authenticate the file.

The same pole and small-gain computations are used at design time,
during training, and at export. During training, a projection clamps
each section's denominator into the Jury stability triangle shrunk by
a margin ($|a_2| \leq 1{-}m$, $|a_1| \leq 1{+}a_2{-}m$, with
$m{=}10^{-3}$ in our experiments; numerators are unchanged) after every
optimiser step. The margin is larger than the verdict band because it
has to hold until the next step as well as across the cast, and the
forward pass only sees iterates inside the stable region. The file
contains everything the real-time engine needs, so the trained graph
and the graph that runs on the audio thread are the same object
(Fig.~\ref{fig:loop}).

Checkpoints trained in FLAMO and in TorchFX's differentiable module
can be imported into the same pipeline. The importer replays each
library's parameter-to-coefficient map in double precision and applies
the same stability check as for a native graph; in CI, fifteen
checkpoints from the two libraries reproduce their realised
coefficients to within float32 rounding ($<5{\times}10^{-5}$ per
coefficient). It handles SISO filters only.

\subsection{The real-time engine}
\label{sec:arch:rt}
The real-time engine runs frozen graphs (cascades, delays, echo,
reverb, FIR) block by block with pre-allocated state, and receives
parameter changes over a lock-free queue. Preparing a graph for a
maximum block size is the only step that allocates: it sizes each
node's scratch buffer once, and processing then works in place. The
queue is a single-producer, single-consumer ring of plain copyable
values, so a coefficient update travels as a fixed-size array of
sections instead of a heap-allocated vector. SOS sections run in
transposed direct form II on every path, so the certified coefficients
describe the same recurrence everywhere. To avoid clicks, new
coefficients are designed off the audio thread and faded in with a
short linear crossfade. We use a linear fade instead of an equal-power
one because both branches process the same input: their outputs are
largely correlated, and gains that sum to one approximately keep the
level.

Besides measuring latency, we test the audio thread with an
instrumented allocator. The test suite installs a global allocator
that counts allocations (not frees) inside the audio-thread section.
The tests cover a cascade with gain automation sent from a second
thread and a nested series--parallel graph, and a control test plants
an allocation to confirm that the counter fires. This shows that the
tested paths do not allocate, which is weaker than a proof. The queue
is lock-free by design, but locks are not instrumented.

\medskip\noindent\textbf{GPU execution.} The batched forward kernels
and the operator interfaces used by their adjoints are written once in
a portable kernel language and run directly on tensors kept on the
device. We measure this path on CUDA hardware only. The adjoint kernels
pass the same gradient checks as the CPU reference in CI; other
backends are not yet validated. Sec.~\ref{sec:eval-thr} measures when
the GPU helps.

\medskip\noindent\textbf{Deployment.} Fluxion ships as Rust crates
with a SoX-style command-line interface, as Python wheels that
exchange arrays through DLPack (zero-copy with NumPy, PyTorch, and
JAX), and as experimental C-ABI and WebAssembly builds. The stripped
release CLI is 3.2\,MB on x86-64 and 2.8\,MB cross-compiled for ARM64,
and it processes 10\,s of audio in 20\,ms
(Tab.~\ref{tab:footprint}); a comparable TorchFX deployment first
loads CPython and PyTorch.

\section{Evaluation}
\label{sec:eval}

\subsection{Throughput}
\label{sec:eval-thr}

\subsubsection{Setup}
\label{sec:eval-thr:setup}
We ran all cells of Tab.~\ref{tab:throughput} on one consumer
workstation, an Intel i9-10900KF (20 threads, 32\,GB RAM) with an
NVIDIA RTX~3070 (8\,GB; driver 580.159, CUDA 12.4) under Linux 6.8.
The baselines run on Python 3.10.12 with PyTorch 2.10.0+cu128. The
edge device of Sec.~\ref{sec:eval-rt} is a Raspberry Pi~5
(4$\times$Cortex-A76 at 2.4\,GHz, 8\,GB, 64-bit Linux 6.18) running
Fluxion cross-compiled on the workstation. All measurements use
float32 and report the median of five timed repetitions after two
warm-up runs (seven after one warm-up for the command-line
comparison), and CUDA timings are bracketed by device synchronisation.
Multi-thread cells use each library's default thread count, 10 for
PyTorch (one per physical core) and 20 for Fluxion; with 10 threads,
Fluxion's cells change by $-2$\% to $+13$\% and no ranking changes.
We report GPU
results in two regimes: \emph{resident}, with tensors kept on the
device as in training, and \emph{one-shot}, where they are uploaded,
filtered, and downloaded as in a single utility call. Exact-recurrence
systems and frequency-sampling references share the table, but our
comparisons refer to the exact-recurrence rows.

\textbf{Systems compared.} Each baseline runs on the path it was
designed for. TorchFX~\cite{spanio2025torchfx}, our earlier system, is
the strongest CPU baseline and has native CUDA kernels. torchaudio's
\texttt{lfilter} is the exact recurrence used in most PyTorch audio
work; it has no SOS primitive, so a cascade costs $K$ kernel
dispatches. GRAFX~\cite{lee_grafx_2024} is
measured with its exact and its frequency-sampling backends.
FLAMO~\cite{flamo} has no time-domain IIR, so its cells report its
frequency-sampling computation and are marked as approximate. SciPy's
second-order-section routine is the single-threaded C reference, and
SoX is the command-line reference. We exclude nnAudio,
ADAC~\cite{adac}, and general-purpose inference runtimes because none
of them runs a sample-recurrent filter (Sec.~\ref{sec:related}), and
compare with anira~\cite{ackvaschulz2024anira} separately
(Sec.~\ref{sec:eval-rt:safety}) on a 1024-tap FIR that both systems
can run.

\subsubsection{Correctness}
\label{sec:eval-thr:correctness}
Before timing, we compared Fluxion's outputs with SciPy. Across the
filter families (Butterworth, Chebyshev~I/II, the RBJ biquads, FIR),
the maximum absolute deviation from SciPy's reference designs on the
same 4000-sample float32 noise signal is $2.7{\times}10^{-5}$ for the
RBJ biquads and at most $5.7{\times}10^{-6}$ for the other families. A
CI test checks that the streaming executor produces the same samples
as whole-signal processing.

\begin{figure}[t]
  \centering
  \includegraphics[width=\linewidth]{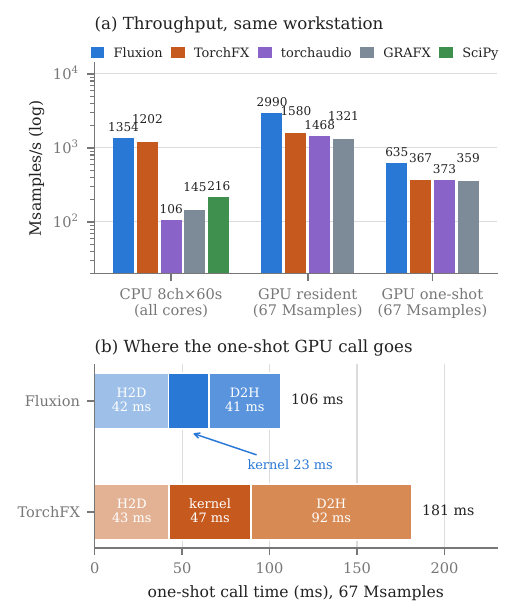}
  \caption{Throughput on one workstation (RTX 3070 + i9).
    (a)~Cross-system medians, with separate resident and one-shot GPU
    regimes. (b)~One-shot decomposition at 67\,Msamples, with device
    synchronisation between stages. With pinned staging, Fluxion's
    transfers reach 6.4\,GB/s and the kernel takes 23\,ms; TorchFX has
    a slower kernel and a slower download to pageable memory.}
  \label{fig:throughput}
\end{figure}

\subsubsection{Results}
\label{sec:eval-thr:results}
\textbf{CPU.} With multiple threads, Fluxion is the fastest
exact-recurrence system on the three CPU workloads: 1354\,Msamples/s
against TorchFX's 1202 on the 8-channel workload, 109 against SciPy's
92 and TorchFX's 63 on the mono 16th-order cascade, and 2603 against
1893 on the 64-signal batch. With one thread, Fluxion is faster on the
batch (897 vs.\ 364) and on the cascade, and TorchFX is faster on the
8-channel workload (407 vs.\ 288). With multiple threads, Fluxion is
9--41$\times$ faster than torchaudio and GRAFX and 1.1--1.7$\times$
faster than TorchFX. Its CPU kernels parallelise over the two
independent dimensions of an IIR workload~\cite{spanio2025torchfx},
sections within a cascade and signals within a batch. The cascade is
fused per sample, and batches go through an AVX2
8$\times$8-transpose path with padded scratch tiling to avoid cache
conflicts; both paths give bit-identical results to the scalar
reference.

\textbf{FIR.} A 2047-tap FIR over 10\,s of audio takes 2.7\,ms with
Fluxion's FFT-based block convolution, 4.8\,ms with SciPy's
overlap-add convolution, and 44.5\,ms with torchaudio's FFT
convolution.

\textbf{Command line.} On a SoX task (a 60\,s stereo WAV through the
same order-16 cascade, from decoding to encoding), the two tools take
the same time (Fluxion 176.2\,ms, SoX 174.9\,ms, medians), and SoX uses
less memory (3.8 vs.\ 5.4\,MB peak resident set).

\textbf{GPU, resident.} With tensors on the device, Fluxion's CUDA
path processes 2990\,Msamples/s, against 1580 for TorchFX's CUDA path,
1468 for torchaudio, and 1321 for GRAFX, and its output matches the
CPU executor to within $10^{-4}$. On the same workload, Fluxion's
multi-threaded CPU path reaches 2662\,Msamples/s, so the GPU adds only
12\%. Among the approximations, GRAFX's FIR-convolution backend does
not gain from the GPU (1199), while FLAMO's spectral multiplication
reaches 8964, much faster than any exact recurrence.

\textbf{GPU, one-shot.} In the one-shot regime, transfers dominate and
the kernel takes 23\,ms of the round trip. With pinned staging for
large uploads, transfers run at 6.4\,/\,6.5\,GB/s and the call takes
106\,ms when Fluxion owns the input buffer
(Fig.~\ref{fig:throughput}b), against about 180\,ms for the
torch-hosted exact-recurrence systems, which download to pageable
memory. With a borrowed slice, as passed by the Python binding, an
extra host copy raises the call to 285\,ms. At this size the CPU is
the better choice for one-shot calls.

\begin{table}[t]
  \caption{Cross-system throughput (Msamples/s, medians, float32;
    single thread in parentheses), co-run on the workstation of
    Sec.~\ref{sec:eval-thr:setup}. Workloads: batch = 64$\times$524k
    (3 sections), 8ch = 8$\times$60\,s (order 4), mono K8 =
    1$\times$60\,s (order 16); GPU = 16384$\times$4096 (67\,Msamples,
    order 8), resident / one-shot (Fluxion: owned input buffer).
    $^{\dagger}$frequency-sampling approximation, not the exact
    recurrence; $^{\ast}$single-threaded library; FLAMO's batch cell
    does not apply. Bold = best exact-recurrence system per column
    (multi-thread).}
  \label{tab:throughput}
  \centering
  \scriptsize
  \setlength{\tabcolsep}{2.5pt}
  \begin{tabular}{@{}lccccc@{}}
    \toprule
    System & CPU batch & CPU 8ch & CPU mono K8 & GPU res. & GPU 1-shot \\
    \midrule
    \textbf{Fluxion} & \textbf{2603} (897) & \textbf{1354} (288) & \textbf{109} (108) & \textbf{2990} & \textbf{635} \\
    TorchFX & 1893 (364) & 1202 (407) & 63 (60) & 1580 & 367 \\
    torchaudio & 64 (20) & 106 (32) & 9 (8) & 1468 & 373 \\
    GRAFX (\texttt{lfilter}) & 121 (36) & 145 (55) & 9 (8) & 1321 & 359 \\
    GRAFX (\texttt{fsm})$^{\dagger}$ & 30 (19) & 21 (14) & 5 (4) & 1199 & 353 \\
    FLAMO$^{\dagger}$ & --- & 102 & 14 & 8964 & 478 \\
    SciPy \texttt{sosfilt}$^{\ast}$ & 208 & 216 & 92 & --- & --- \\
    \bottomrule
  \end{tabular}
\end{table}

\subsection{Differentiable DSP}
\label{sec:eval-diff}

\begin{figure}[t]
  \centering
  \includegraphics[width=\linewidth]{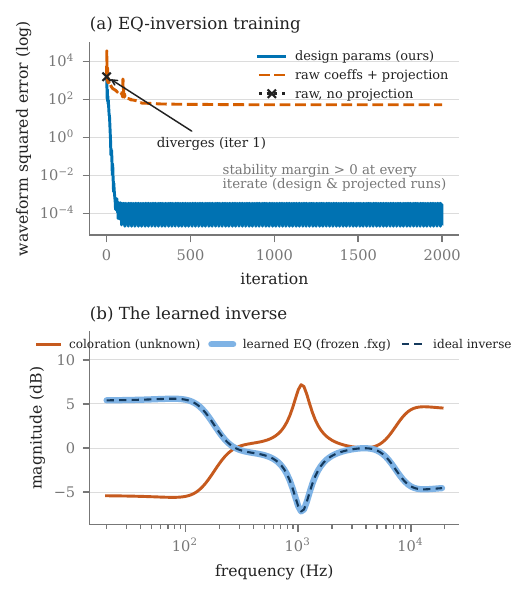}
  \caption{Coloration inversion with a parametric EQ. (a)~Training loss
    for the three parameterisations. (b)~Magnitude response of the
    frozen artifact trained on design parameters, the file that the
    real-time engine plays.}
  \label{fig:loop}
\end{figure}

\subsubsection{Coloration inversion}
\label{sec:eval-diff:task}
A fixed coloration filter, a three-section chain of shelving and
peaking filters, colours a signal, and a parametric EQ has to invert
it. The trainer sees the coloured and the original signals but not the
filter, and minimises the time-domain squared error between the
equalised signal and the original. The task is simple, but the ideal
inverse is known, so we can measure the residual.

We compare three parameterisations: the EQ's nine design parameters
(three frequencies, gains, and Q factors), as proposed in
Sec.~\ref{sec:arch:diff}; the fifteen raw biquad coefficients with the
stability projection; and the raw coefficients without projection.
Each run takes 2000 iRprop$^-$ steps~\cite{igel2000rprop} on the same
8192-sample pink-weighted noise at 48\,kHz, using the whole signal at
every step. iRprop$^-$ moves each parameter by the sign of its
gradient, so we do not need a single learning rate across hertz,
decibels, and Q; a step grows by a factor of 1.2 while the sign holds
and halves when it flips. Because the optimiser uses only the sign,
this task does not test gradient accuracy, which
Sec.~\ref{sec:eval-diff:verify} covers.

With design parameters, the residual is 0.0014\,dB RMS (0.004\,dB
worst case on a 120-point log-frequency grid), and the learned
response overlays the ideal inverse in Fig.~\ref{fig:loop}b. The
stability margin stays positive by construction. After about 200
iterations the loss stops decreasing and oscillates between
$2{\times}10^{-5}$ and $3.4{\times}10^{-4}$ (Fig.~\ref{fig:loop}a), an
effect of the optimiser's step adaptation. The raw-coefficient run
with the projection stays stable at every iterate but stalls at a loss
of 52, with a residual of 0.83\,dB RMS (2.3\,dB worst case); with this
optimiser, the nine design parameters are easier to fit than the
fifteen coefficients. Without the projection, the first update already
leaves the stable region and the forward pass diverges. The EQ trained
on design parameters is then frozen and played by the real-time
engine, and Fig.~\ref{fig:loop}b shows the response of that file.

\medskip\noindent\textbf{Comparison with FLAMO.} On the same task, a
three-section generic SVF trained as in FLAMO~\cite{flamo}
(frequency-sampled magnitude response, dB-MSE, Adam, 4000 iterations
in 12.3\,s) reaches a residual of 0.44\,dB RMS (1.11\,dB worst case).
This is better than Fluxion's raw-coefficient run and worse than its
design-parameter run, so here the parameterisation matters more than
the way gradients are computed. The Fluxion runs also give a stability
verdict at every iterate and an artifact that the real-time engine
plays, whereas FLAMO optimises the magnitude response only and deploys
through a separate compiler~\cite{adac}.

\subsubsection{Training cost}
\label{sec:eval-diff:throughput}
Tab.~\ref{tab:trainthr} reports one forward and backward step of an
SOS cascade with five methods: Fluxion's analytic adjoints, TorchFX's
differentiable module, torchaudio's built-in differentiable recursion,
GRAFX's autograd over the same backend, and FLAMO's frequency-sampling
step. FLAMO's step computes one FFT-domain multiplication per section,
so its cost barely depends on $K$ (3--5\,ms), and its gradient
approximates the exact recursion that the other rows compute. The
Fluxion column uses the CPU adjoint path and processes the signals of
the batch one at a time. On CPU it is slower than torchaudio and GRAFX
in every measured cell (1.4--3.2$\times$), and faster than TorchFX only
at $T{=}16$k with $K \geq 4$. The caption gives the cost of an
unrolled recursion.

\medskip\noindent\textbf{GPU training.} The analytic VJP kernels also
run on the GPU with data kept on the device, in two configurations.
Gradients through a frozen cascade to its input, as needed for a DSP
layer inside a larger model, cost 40--144\,ms per step at $T{=}16$k
and 137--714\,ms at $T{=}64$k for $K \in \{1,4,8\}$ (batch 16). A
trainable single section costs 32\,ms at $T{=}16$k and 116\,ms at
$T{=}64$k, which is slower than Fluxion's CPU adjoint (11 and 46\,ms)
because the current entry point launches one row at a time and
synchronises the device after each row. A batched trainable cascade is
future work. For now, the CPU path trains faster at batch 16, and the
GPU helps only in the resident forward path of
Sec.~\ref{sec:eval-thr}.

\begin{table}[t]
  \caption{One training step (forward and backward, squared-error loss;
    ms, median) of a $K$-section cascade over a batch of 16 signals of
    $T$ samples. Fluxion: analytic adjoints on the CPU, one signal at a
    time. TorchFX \texttt{ddsp}: analytic adjoints over the native
    kernel (CPU/CUDA). torchaudio \texttt{lfilter}: its built-in
    differentiable path (CPU/CUDA). GRAFX: autograd over its exact
    \texttt{lfilter} backend (CPU). FLAMO$^{\dagger}$:
    frequency-sampling step (magnitude-response MSE at $nfft{=}T$),
    whose cost barely depends on $K$. An unrolled recurrence takes
    408\,ms at $K{=}1, T{=}4$k, about
    153$\times$ the analytic path per sample.}
  \label{tab:trainthr}
  \centering
  \scriptsize
  \setlength{\tabcolsep}{2.5pt}
  \begin{tabular}{@{}ccccccc@{}}
    \toprule
    $K$ & $T$ & Fluxion & TorchFX ddsp & torchaudio & GRAFX & FLAMO$^{\dagger}$ \\
    \midrule
    1 & 16k & 11 & 10 / 15 & 4 / 7 & 3 & 3 \\
    1 & 64k & 46 & 22 / 16 & 17 / 30 & --- & --- \\
    4 & 16k & 26 & 38 / 63 & 15 / 26 & 13 & 5 \\
    4 & 64k & 114 & 82 / 63 & 75 / 113 & --- & --- \\
    8 & 16k & 48 & 73 / 126 & 31 / 52 & 26 & 5 \\
    8 & 64k & 207 & 162 / 126 & 149 / 221 & --- & --- \\
    \bottomrule
  \end{tabular}
\end{table}

\subsubsection{Gradient checks}
\label{sec:eval-diff:verify}
Finite differences are unreliable through a deep recursive filter,
especially in single precision, so we check gradients at two levels.
Each operator's VJP (input, coefficient, design-parameter, FIR, and
delay gradients) is compared with finite differences in isolation,
where the comparison is well conditioned. The gradient of a mixed
series--parallel graph is compared with an independently hand-composed
adjoint to within $10^{-4}$, and the GPU backward kernels pass the
same checks on the device. All checks run in continuous integration.

\subsection{Real-Time and Edge}
\label{sec:eval-rt}

\subsubsection{Real-time behaviour}
\label{sec:eval-rt:safety}
\begin{figure}[t]
  \centering
  \includegraphics[width=\linewidth]{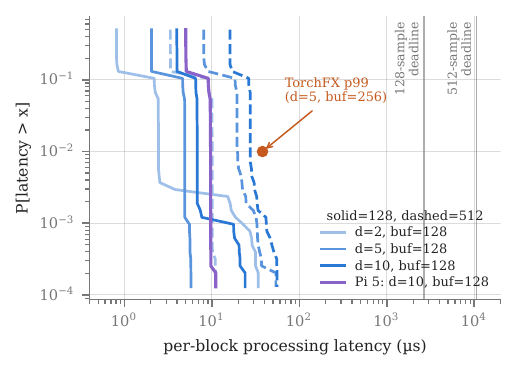}
  \caption{Per-block processing time (unpaced) at 128 and 512 samples.
    The violet curve is the same code on a Raspberry Pi 5; the orange
    point is TorchFX's p99 from our earlier study.}
  \label{fig:ccdf}
\end{figure}

The allocation tests of Sec.~\ref{sec:arch:rt} run in CI and find no
allocations on the tested audio-thread paths.
Fig.~\ref{fig:ccdf} shows the per-block processing
time at cascade depths $d \in \{2, 5, 10\}$ and buffer sizes
$\{128, 256, 512\}$ (deadlines of 2.7--10.7\,ms at 48\,kHz), with
coefficient crossfades running throughout. Blocks are processed one
after another, without pacing or real-time priority, so the figure
shows computation cost and not operating-system scheduling delays.

In every configuration, the p99 is at most 0.26\% of the deadline
(6.8\,$\mu$s at depth 10 with 128-sample buffers). The slowest of the
73{,}728 blocks took 872\,$\mu$s (depth 5, 512 samples, 8\% of that
deadline); in the other configurations, the slowest block took less
than 66\,$\mu$s. Our earlier study measured a TorchFX p99 of
0.71--0.74\% of the deadline on the same machine at 256 samples, the
smallest buffer it used.

\medskip\noindent\textbf{Comparison with anira.}
anira~\cite{ackvaschulz2024anira} runs neural inference outside the
audio callback, on a thread pool. By default the callback returns an
earlier result, which adds one buffer of latency (2.7--10.7\,ms here).
With a blocking ratio $W > 0$, the callback waits up to $W$ buffer
periods for the current result, and anira adds no latency when that
wait covers the declared worst-case inference time. We ran a 1024-tap
FIR with the same taps in both systems: anira runs it as an ONNX
convolution (ONNX Runtime, $W{=}0.75$, which adds no latency at the
three buffer sizes), and Fluxion runs its real-time FIR node. Both run
unpaced and without real-time priority, over at least 5{,}000 blocks
per configuration. At 128/256/512 samples, Fluxion processes a block
in 25/51/101\,$\mu$s (median; p99 27/68/124\,$\mu$s), and anira's
callback returns in 81/91/150\,$\mu$s (p99 173/177/227\,$\mu$s); no
anira block reached the wait limit. The anira process has a 363\,MB
resident set and the Fluxion process 2.7\,MB. anira can run arbitrary
neural models and Fluxion cannot, so this comparison applies only to
DSP workloads.

\begin{table}[t]
  \caption{Deployment footprint. The Fluxion rows process 10\,s of
    audio (16th-order cascade and reverb, file to file) on the
    workstation and on a Raspberry Pi 5 (binary cross-compiled,
    zig-linked); the Python row is the cost of
    \texttt{import torch, torchfx} alone.}
  \label{tab:footprint}
  \centering
  \footnotesize
  \begin{tabular}{@{}lccc@{}}
    \toprule
    Stack & On disk & Peak RSS & Wall \\
    \midrule
    Fluxion CLI (x86-64, stripped) & 3.2\,MB & 11\,MB & 20\,ms \\
    Fluxion CLI (Pi 5, aarch64) & 2.8\,MB & 11\,MB & 46\,ms \\
    CPython + torch + torchfx (import only) & 1.7\,GB & 625\,MB & 4.3\,s \\
    \bottomrule
  \end{tabular}
\end{table}

\subsubsection{Footprint}
\label{sec:eval-rt:edge}
Tab.~\ref{tab:footprint} compares the two stacks on the same task.
Fluxion's ARM64 release binary is a 2.8\,MB executable without Python
or PyTorch dependencies, while a TorchFX deployment loads
CPython and PyTorch before it processes the first sample.

\medskip\noindent\textbf{Raspberry Pi 5.} The cross-compiled binary
runs the 16th-order cascade and reverb chain at 217$\times$ real time
with an 11\,MB resident set on a Pi~5 (4$\times$Cortex-A76, 8\,GB).
With the unpaced protocol above and coefficient crossfades running,
the p99 is at most 0.36\% of the deadline over all depths and buffer
sizes, and the slowest block took 41\,$\mu$s (Fig.~\ref{fig:ccdf}).

\medskip\noindent\textbf{Hot-swap on the device.} We ran the train,
certify, push, and crossfade sequence of
Sec.~\ref{sec:discussion:research} end to end. The EQ of
Sec.~\ref{sec:eval-diff}, trained on the workstation, was copied as a
frozen \texttt{.fxg} file over the LAN to the Pi~5 while the engine
streamed 128-sample blocks. The device recomputes the certificate when
the file arrives and then crossfades the new filter in through the
lock-free command queue. Loading, certification, freezing, and
enqueueing took 27.5\,$\mu$s on the device. The whole update, from the
start of the copy to the engine's acknowledgement, took 1.24\,s,
mostly for the network copy and polling. During the 43\,ms crossfade,
the largest sample-to-sample step of the output was 0.036, against
0.029 in steady state. In the same run, a file edited after
certification to make it unstable (echo feedback changed from 0.9 to
1.5) was rejected on the device and never reached the audio thread.

\begin{figure}[t]
  \centering
  \includegraphics[width=0.6\linewidth]{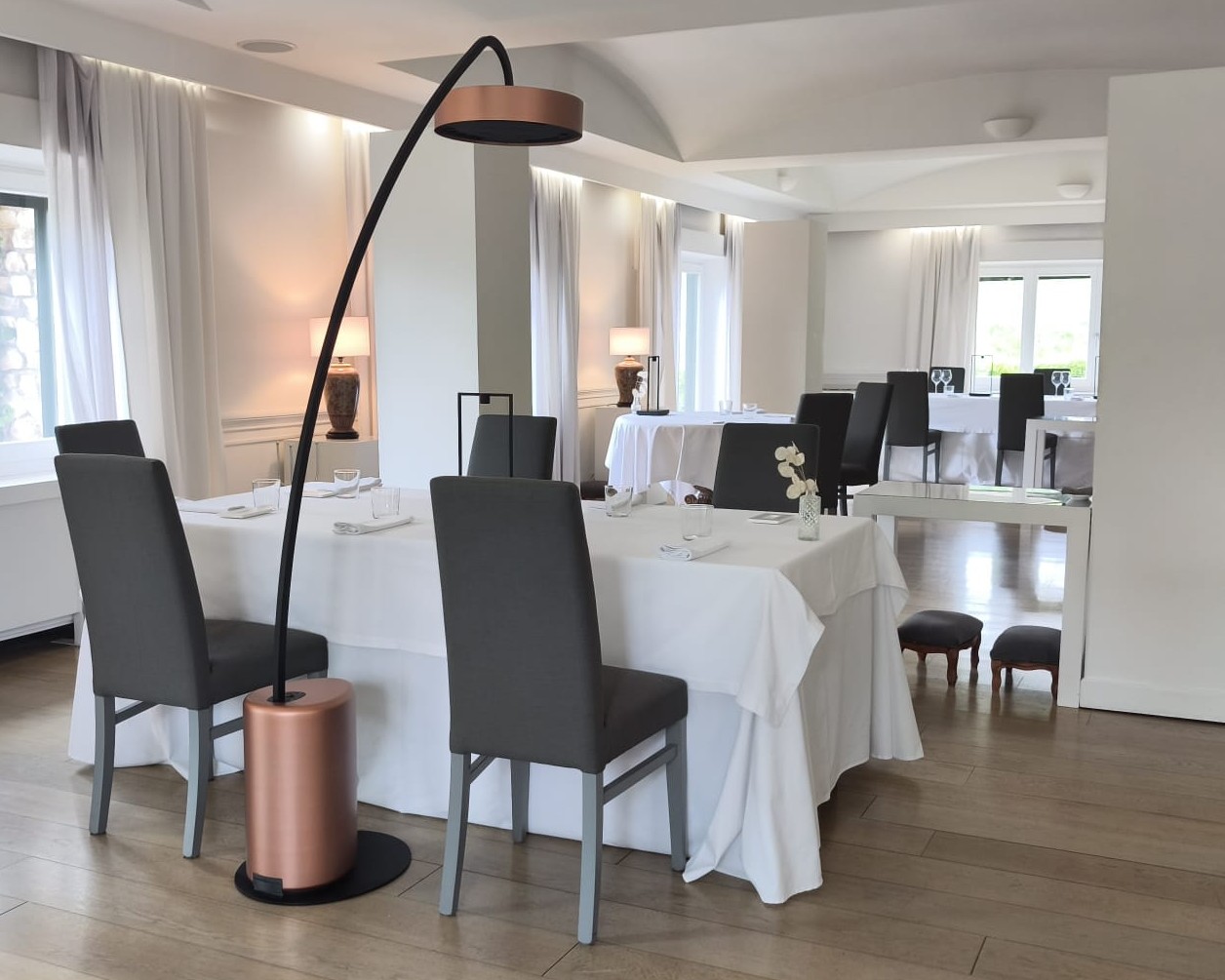}
  \caption{The SoundFood soundlamp, a 7.1 smart speaker that plays
    AI-generated music matched to each dish in restaurants. Its
    equalisation and directional FIR chain runs in real time on a
    Raspberry~Pi-class core (Sec.~\ref{sec:eval-rt:soundfood}).}
  \label{fig:soundlamp}
\end{figure}

\subsubsection{Case study: SoundFood}
\label{sec:eval-rt:soundfood}
SoundFood (\url{https://www.soundfood.it/}) develops a smart speaker
for restaurants (Fig.~\ref{fig:soundlamp}) that plays AI-generated
music matched to each dish~\cite{spanio2024emotionally,spanio2025multimodal},
an application of ``sonic seasoning'', the use of sound to shape taste
perception~\cite{spence2008multisensory}. Because tables start their
dishes at different times, the device keeps each table's music within
a local ``sound bubble''. The production chain upmixes a mono track to
eight channels (7.1). Each channel runs six frozen biquad sections (two
order-4 Butterworth band limits, one shelving and one peaking section),
its own 1024-tap directional FIR, and a level trim, and a separate
low-frequency feed uses an order-2 Linkwitz-Riley crossover and an
order-4 Butterworth. We report the structure of the chain; its tuning
(cutoffs, gains, and taps) belongs to SoundFood.

The system started as an offline batch service (FastAPI over TorchFX)
that pre-rendered and stored every track eight times (5.3\,MB of mono
source becomes 42.3\,MB per minute of 7.1 output), holding the whole
file in memory (0.6\,GB peak for a one-minute track). On the device,
the newest TorchFX wheel for ARM runs the chain at $0.72\times$ real
time (83.8\,s for a 60\,s track, 661\,MB peak, a 968\,MB installation,
and a 2.1\,s import), and TorchFX's compiled kernels are built for x86
only.

Fluxion loads the frozen biquad rows and FIR taps of the production
chain directly. The largest output deviation from the production
service is $2.0\times10^{-4}$ full scale on one channel and at most
$9.6\times10^{-5}$ on the others; the reference output is 16-bit PCM
(step $3.1\times10^{-5}$). On the Pi~5, Fluxion runs the chain as
eight separate certified graphs. Paced at the audio rate, the p99
block latency is at most 33\% of the deadline at buffer sizes of 128,
512, and 1024 samples, with no missed deadlines over 57{,}000 blocks,
31\% of one core, and a 13.5\,MB resident set from a 468\,KB harness
binary. Offline, the chain runs at $4.2\times$ real time on the
device. The equalisation can therefore run on the device instead of
being pre-rendered, and per-dish retuning could use the hot-swap
described above.

\section{Discussion}
\label{sec:discussion}

\subsection{Summary of results}
Among the exact-recurrence systems, Fluxion has the highest throughput
in the resident GPU regime and in most CPU cells (with one thread,
TorchFX is faster on the 8-channel workload), and it takes the same
time as SoX on the file-to-file task. Its CPU training step is slower
than those of torchaudio and GRAFX, and its GPU training path is still
slower than its CPU path. The deployment footprint is about 500$\times$
smaller on disk and 57$\times$ smaller in memory than the Python
stack, and none of the measured baselines covers training,
certification, freezing, and playback in one library. Fluxion is a DSP
runtime and does not offer torchaudio's range of transforms.

\subsection{Research directions}
\label{sec:discussion:research}
Design-parameter gradients turn filter design into an optimisation
over interpretable parameters that keep the filter stable. This could
be used to study perceptual losses, style transfer between EQ curves,
and regularisation that is awkward to express over raw coefficients.
Versioned frozen artifacts, a stability check at export, and
crossfaded coefficient swaps let edge devices replace retrained DSP
while streaming, with stability checked on the artifact itself.
Managing fleets of devices, versioning, and staged rollout of DSP
updates are open problems.

A frequency-sampling step like FLAMO's, whose cost barely depends on
depth (Tab.~\ref{tab:trainthr}), could be added to Fluxion for coarse
training, followed by certification and fine-tuning with the exact
adjoints on the same graph. An adjoint for the feedback operator,
together with a loop-aware stability check during training, would
make learnable reverberators and feedback delay networks possible;
keeping these stable during training is hard, and the stability
verdicts of Sec.~\ref{sec:arch:freeze} could be a starting point.

\subsection{Limitations}
GPU execution is validated on CUDA only. The kernel layer also targets
Metal, ROCm, and WebGPU, but we have not tested it on that hardware.
GPU training is measured only for the single trainable section and for
gradients through frozen cascades (Sec.~\ref{sec:eval-diff}), and
trainable multi-section cascades run on the CPU adjoint path. The
feedback operator executes but is neither differentiable nor lowered
to the real-time engine. The kernels do not force flush-to-zero for
denormals, which matters for decaying tails near digital silence.
Fluxion has no neural operators of its own. The allocation claim
relies on instrumented tests of specific paths, not on static analysis
of all reachable paths. Frees and locks are not instrumented, and the
property holds per process, not at the operating-system level.

\subsection{Threats to validity}
All numbers come from one workstation and one edge board, and other
hardware could change the order of close cells (the smallest CPU
margin is 13\%). Each baseline runs on its own best path under one
float32 protocol, and we state the remaining protocol differences
where the numbers appear: frequency-sampling approximations, SoX's
file-to-file pipeline, the callback-to-return measurement of our
earlier study, the FIR-only comparison with anira, and the default
thread counts. Timings are medians of five repetitions after warm-up,
and Tab.~\ref{tab:throughput} and Tab.~\ref{tab:trainthr} report them
without dispersion, so the closest comparisons (the 13\% CPU cell and
the command-line comparison) should be read as ties. The latency
distributions of Sec.~\ref{sec:eval-rt} were measured without pacing
or real-time priority.

\section{Conclusions and Future Work}
\label{sec:conclusion}

Fluxion implements the DSP-specific numerical code (closed-form design,
forward kernels, analytic adjoints, and stability checks) and uses
existing libraries for autodiff, GPU code generation, and array
interchange. One graph algebra is lowered to a training engine and to
a real-time engine whose audio thread we test for allocations, and the
freeze step carries a checked artifact from training to the edge
device, including checkpoints imported from FLAMO and TorchFX.

Future work includes testing the Metal, ROCm, and WebGPU kernels on
hardware, making the feedback operator differentiable and usable in
real time with a loop-aware stability check during training, adding
frequency-sampling gradients as a training option, and defining
versioning and rollout policies for DSP updates on many devices. We do
not plan fp16 or bf16 execution, because recursive filters are
sensitive to coefficient quantisation at half precision.

\section*{Disclosure of AI Use}
We used large language model assistants when drafting and revising
parts of this manuscript and as coding assistants while implementing
Fluxion. They were not used to produce or to interpret any
measurement. The authors derived and checked the adjoints, designed
and ran every experiment, verified the code and the numbers reported
here, and take full responsibility for them.

\section*{Acknowledgment}
This work was funded by the European Union - NextGenerationEU, under the National Recovery and Resilience Plan (PNRR).

\bibliographystyle{IEEEtran}
\bibliography{refs}

\end{document}